\pdfoutput=1

\documentclass[3p,times,procedia]{elsarticle}
\usepackage{nupha_ecrc}

\volume{00}

\firstpage{1}

\journalname{Nuclear Physics A}

\runauth{}

\jid{nupha}

\jnltitlelogo{Nuclear Physics A}

\usepackage{amssymb}

\usepackage[figuresright]{rotating}

\begin{document}

\begin{frontmatter}



\dochead{}

\title{Effects of enhanced bulk viscosity near the QCD critical point}


\author[label1]{Akihiko Monnai}
\author[label2]{Swagato Mukherjee}
\author[label3]{Yi Yin}
\address[label1]{Institut de Physique Th\'{e}orique, CNRS URA 2306, CEA/Saclay, F-91191
Gif-sur-Yvette, France}
\address[label2]{Department of Physics, Brookhaven National Laboratory, Upton, New York
11973-5000}
\address[label3]{Center for Theoretical Physics, Massachusetts Institute of Technology, Cambridge, MA 02139 USA}

\address{}

\begin{abstract}
Search for the conjectured QCD critical point is one of the major scientific goals for the Beam Energy Scan program at RHIC. The growth of the correlation length is a universal feature for systems near criticality, and bulk viscosity exhibits the strongest dependence on the correlation length. We investigate its effects using a relativistic hydrodynamic model to find that rapidity distributions of charged particles and net baryon number are visibly modified if the fireball passes through the vicinity of the QCD critical point on the phase diagram. We also discuss how critical modification of dilepton emission rate may leave imprints on invariant mass spectra.
\end{abstract}


\end{frontmatter}


\section{Introduction}
\label{sec:1}

Exploration of the QCD phase diagram has been one of the most intriguing issues in hadron physics. The discovery of the quark-gluon plasma (QGP) phase at BNL Relativistic Heavy Ion Collider (RHIC) has been the first major experimental step. The collider experiment has brought us unprecedented amount of information about the properties of the quark matter, including the fact that the QGP is a relativistic fluid. CERN Large Hadron Collider (LHC) experiment has added a new insight that the medium remains to be fluid-like at larger temperatures. The properties of the QCD medium at large chemical potential are now a topic of great interest, and a number of experimental projects, including the Beam Energy Scan program at RHIC, are devoted for the search for the QCD critical point.

It is an important task to establish what observables are sensitive to the critical point, and what signals we would have in those observables. 
In this study, we extend a relativistic viscous hydrodynamic model to finite baryon density and argue that bulk viscosity is essential for the heavy-ion phenomenology in the presence of the critical point \cite{Monnai:2016kud} since the dependences of shear viscosity, bulk viscosity and baryon diffusion on the correlation length $\xi$ in the critical region are $\eta \sim \xi^{(4-d)/19}$, $\zeta \sim \xi^{3}$ and $D_B \sim \xi^{-1}$, respectively. Numerical simulations indicate that medium evolution can be visibly modified when fluid elements pass through the $T$-$\mu_B$ region close to the critical point. We also show that rapidity distributions of charged hadrons and net baryon number can be deformed owing to enhanced entropy production and flow convection caused by the critical behavior of bulk viscosity. Finally we discuss the effect of the critical point on dileptons. 

\section{Viscous hydrodynamic model near QCD critical point}
\label{sec:2}

Relativistic hydrodynamic equations of motion \cite{Israel:1979wp} with bulk viscosity are 
\begin{eqnarray}
\partial_\mu T^{\mu \nu} &=& 0, \\
\partial_\mu N_B^\mu &=& 0, \\
u^\mu \partial_\mu \Pi &=& - \frac{1}{\tau_\Pi} (\zeta \partial_\mu u^\mu + \Pi ),
\end{eqnarray}
where the energy-momentum tensor is decomposed as $T^{\mu \nu} = (e+P+\Pi) u^\mu u^\nu + (P+\Pi) g^{\mu \nu}$ and the net baryon number current as $N_B^\mu = n_B u^\mu$ in the Landau frame. Shear viscosity and baryon diffusion are neglected for simplicity. 
$\tau_\Pi$ is bulk viscous relaxation time. The metric is defined as $g^{\mu \nu} = \mathrm{diag}(-,+,+,+)$.

\begin{figure}[tb]
\center
\includegraphics[width=2.1in]{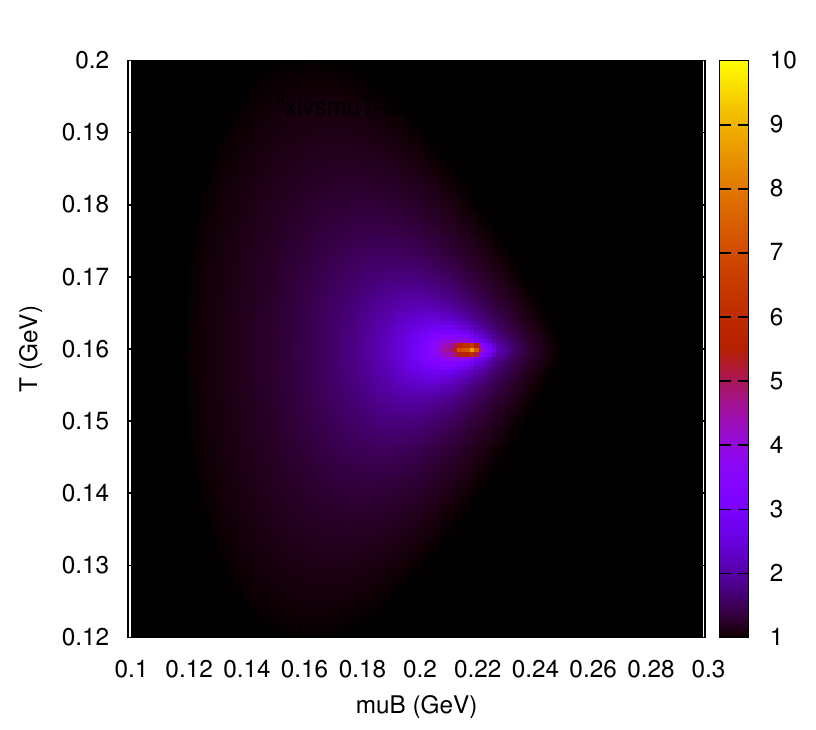}
\includegraphics[width=2.7in]{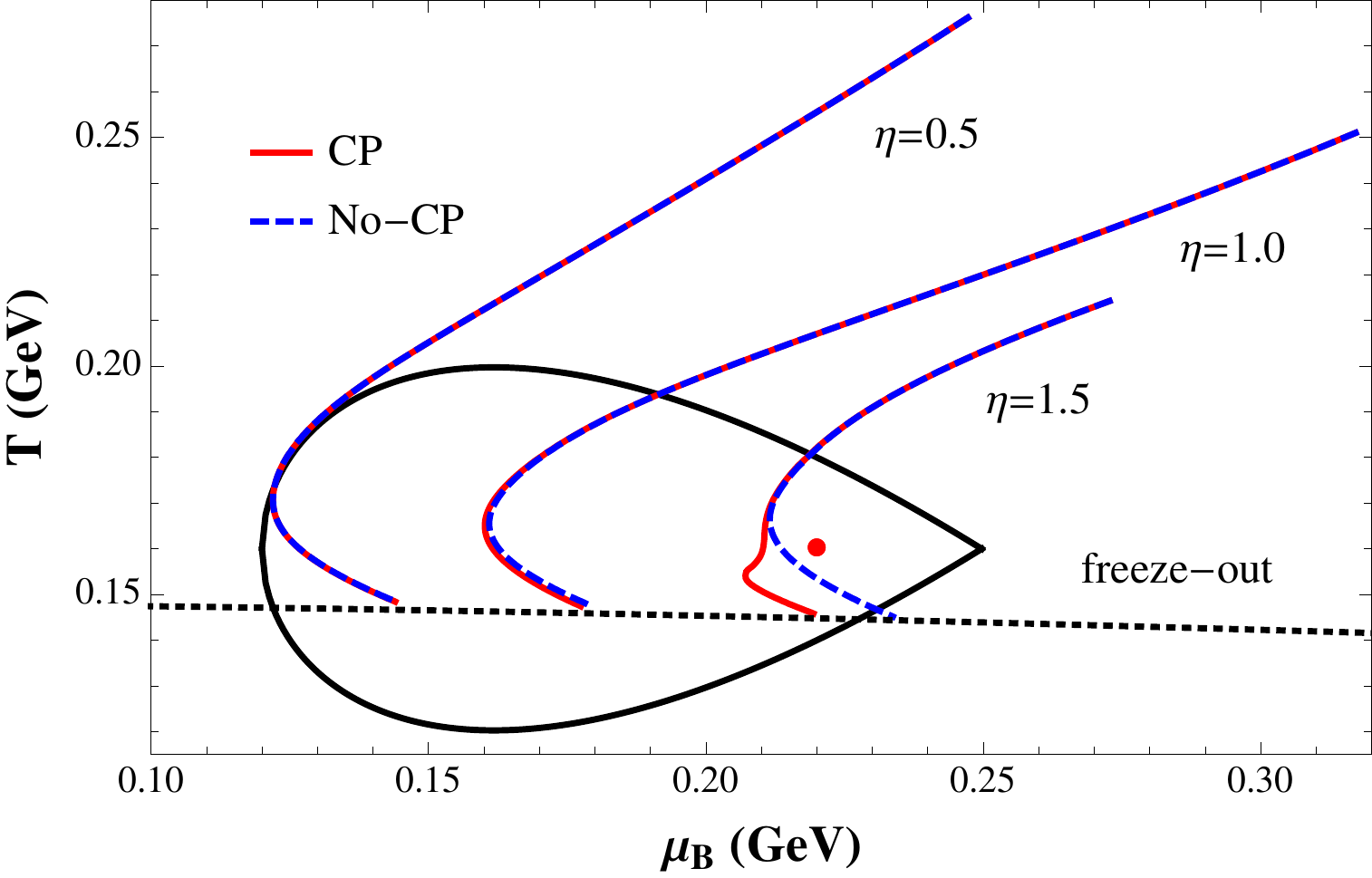}
\caption{(Left) $\xi/\xi_0$ near the critical region. (Right) The trajectories of hydrodynamic medium evolution at given space-time rapidities.}
\label{fig:1}
\end{figure}

We determine the behavior of the transport coefficients in the vicinity of the critical point. Using the insight of the critical point for the liquid-gas transition, which belongs to model H, it is parametrized as $\zeta \sim \xi^{z-\alpha/\nu} \sim \xi^{3}$ at the mean field level \cite{Onuki:1997}. Here $\alpha$ and $\nu$ are equilibrium critical exponents and $z$ is a dynamical critical exponent.
The characteristic damping time of the bulk pressure in the critical region is determined by the relaxation time of the critical mode $\tau_\sigma \sim \xi^z$, which is also related to $\tau_\Pi$. With the supplementally argument that causality is preserved when $\zeta/\tau_\Pi (\epsilon + P)$ remains finite, it is implied that the relaxation time is given as $\tau_\Pi \sim \xi^3$. We thus choose to parametrize the transport coefficients as 
\begin{eqnarray}
\zeta = \zeta_0 \bigg( \frac{\xi}{\xi_0} \bigg)^3 \ \mathrm{and} \ \tau_\Pi = \tau_{\Pi}^{0} \bigg( \frac{\xi}{\xi_0} \bigg)^3 ,
\end{eqnarray}
where 
$\zeta_0 = 2 (1/3- c_s^2) (\epsilon + P)/4\pi T$ and $\tau_{\Pi}^{0} = [18 - (9\ln 3 - \sqrt{3} \pi)]/24\pi T$
are estimated in the holographic approaches \cite{Buchel:2007mf, Natsuume:2007ty}. 
The dependence of the correlation length $\xi$ on $T$ and $\mu_B$ are determined by mapping the critical region of the Ising model using the prescription $(T-T_c)/\Delta T = h/\Delta h$ and $(\mu_B-\mu_B^c)/\Delta \mu_B = - r/\Delta r$ where the Ising variables $h$ and $r$ are rescaled magnetic field and reduced temperature, 
respectively. The maximum of $\xi / \xi_0$ is normalized to $10$. In this study, we take $(T_c,\mu_B^c) = (0.16, 0.22)$ GeV as the location of the critical point and $(\Delta T,\Delta \mu_B) = (0.02, 0.1)$ GeV as the size of the critical region (Fig.~\ref{fig:1} (left)). 
$\mu_B^c$ may be too small, but it should be emphasized that our primary purpose is to demonstrate the critical effects on hydrodynamic evolutions and observables rather than to find the location of the critical point.

For hydrodynamical simulations, we consider a 1+1 dimensional geometry \cite{Monnai:2012jc} to study rapidity dependence of hadronic yields because net baryon number is larger at forward rapidity. The initial conditions of energy and net baryon density profiles are based on the color glass condensate (CGC) extrapolated to 17 GeV. The CGC may not be a good description at those energies but will be suffice for our demonstration. The initial time $\tau_\mathrm{th} = 1.5$ fm/$c$. 
The equation of state at finite density is constructed using the lattice QCD data up to the fourth order of baryon fluctuations \cite{Bazavov:2014pvz,Bazavov:2012jq,Ding:2015fca} and is smoothly matched to that of hadron resonance gas at lower energies. 
In this setup, the conjectured critical point is at space-time rapidity $\eta \sim 1.5$ (Fig.~\ref{fig:1} (right)).

The hadronic distributions are calculated by converting fluid into particles at thermal freeze-out. The freeze-out energy density for the Cooper-Frye formula is $\epsilon_f = 0.25$ GeV/fm$^3$. The effect of bulk viscous distortion of the phase-space distribution is estimated using Grad's moment method \cite{Israel:1979wp}.

\section{Dilepton emission rate with bulk viscosity}
\label{sec:2}

The dilepton emission rate is also affected by the large bulk viscosity. The rate for $a^+ a^- \to l^+ l^-$ \cite{Kajantie:1986dh} is
\begin{eqnarray}
\frac{dN}{d^4x} &=& \int \frac{d^3 p_1}{(2\pi)^3} \frac{d^3 p_2}{(2\pi)^3} f_1 (E_1) f_2(E_2) \sigma(M) v_\mathrm{rel} ,
\end{eqnarray}
where 
$v_\mathrm{rel} = \sqrt{(p_1\cdot p_2)^2 - m_a^4}/E_1 E_2$
and $f_{i}$ $(i=1,2)$ is the distribution function. We consider the quark-antiquark annihilation for the QGP phase and the pion pair annihilation for the hadronic phase for the moment.  
We introduce the thermal mass of partons
$m_\mathrm{q, th}^2 = g^2 ( T^2 + \mu_q^2/\pi^2 )/6$
where $\mu_q = \mu_B/3$.

The invariant mass $M$ spectra in local equilibrium is expressed in the Boltzmann limit as
\begin{eqnarray}
\frac{dN_0}{d^4x d M^2 d^2 p_T dy} &=& \frac{\sigma (M)}{2(2\pi)^5} \frac{M^2}{2} e^{-E/T} \bigg(1-\frac{4 m_a^2}{M^2} \bigg) ,
\end{eqnarray}
where $\sigma (M)$ is the cross section.
The bulk viscous distortion of distribution $\delta f = f - f_0$ can be expressed using Grad's moment method as
$
\delta f^i = - f_0^i [b_i D_\Pi E_i + B_\Pi (m_a^2 - E_i^2) + \tilde{B}_\Pi E_i^2 ] \Pi .
$
The distortion factors $D_\Pi$, $B_\Pi$ and $\tilde{B}_\Pi$ can be computed in kinetic theory. The bulk viscous correction to the emission rate is
\begin{eqnarray}
\frac{d\delta N}{d^4x d M^2 d^2 p_T dy} &=& - \Pi \frac{\sigma (M)}{(2\pi)^5} M e^{-E/T} \bigg(1-\frac{4 m_a^2}{M^2} \bigg) \bigg[ B_\Pi m_a^2 \frac{M}{2} + (\tilde{B}_\Pi - B_\Pi) \frac{M^3}{8} \bigg] . 
\end{eqnarray}
It should be noted that the correction rate is sensitive to the components in the target system. Here we consider pion gas for the hadron phase and parton gas for the QGP phase.

\section{Results}
\label{sec:3}

\begin{figure}[tb]
\center
\includegraphics[width=2.6in]{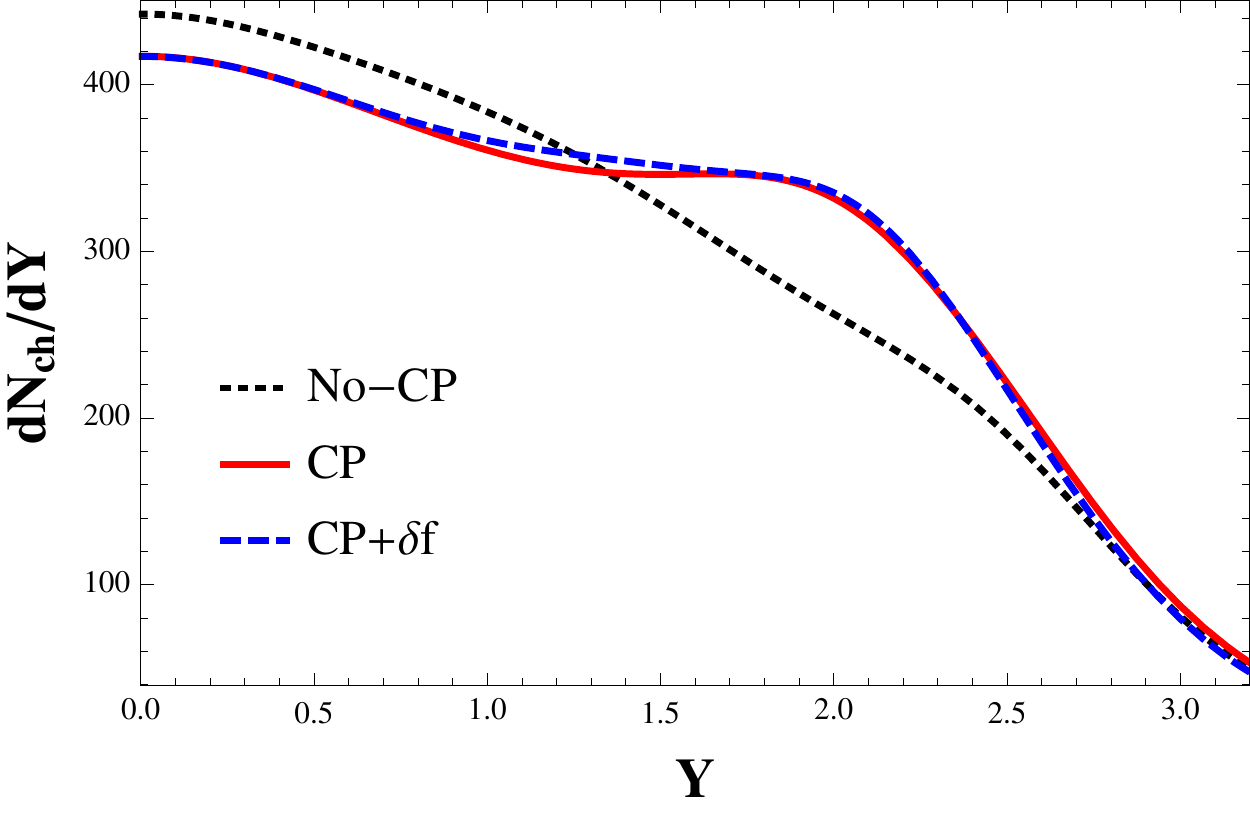}
\includegraphics[width=2.6in]{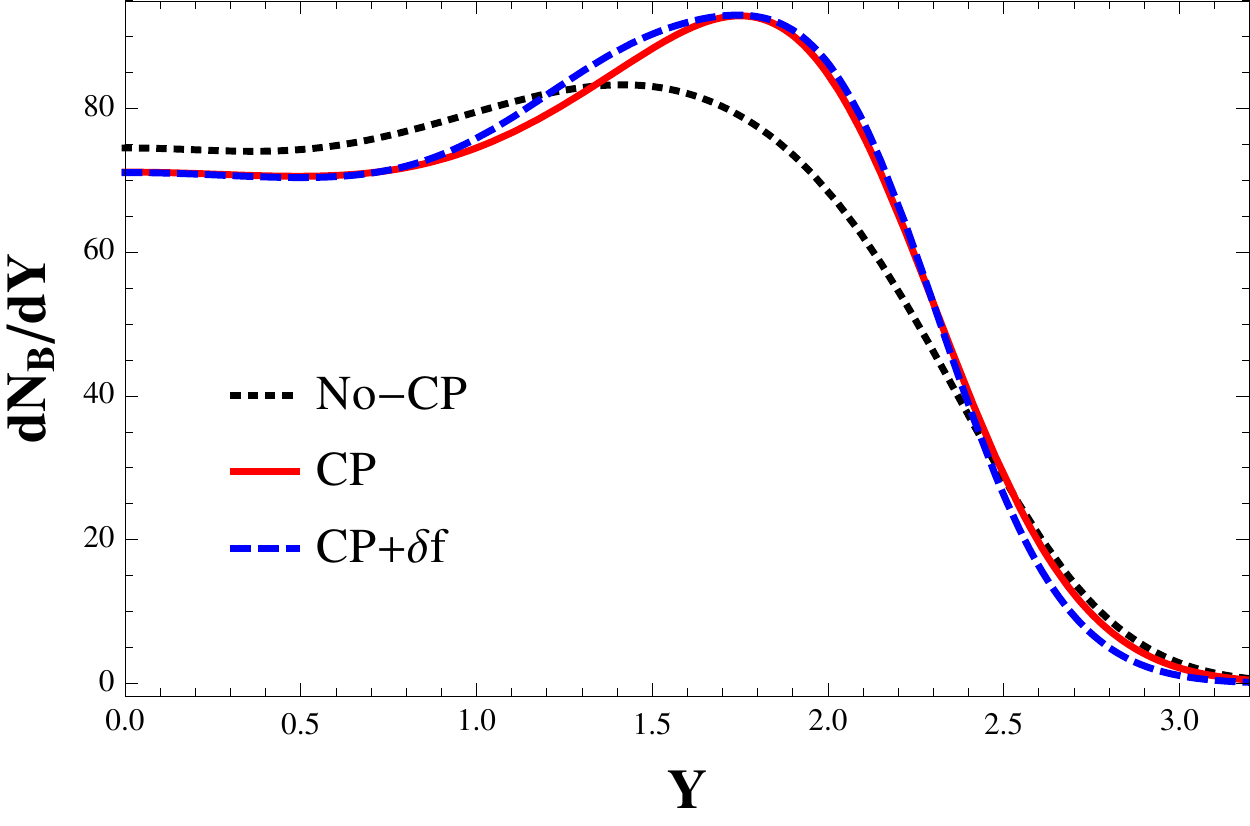}
\caption{(Left) Charged particle rapidity distribution and (right) net baryon rapidity distribution.}
\label{fig:2}
\end{figure}

Figure~\ref{fig:2} shows the rapidity distributions of charged hadrons and net baryon number with and without the conjectured QCD critical point. The charged particle distribution is visibly enhanced at forward rapidity where the medium passes through the critical region. This is caused by two factors: extra entropy production owing to the large viscosity and enhancement of the gradient in the effective pressure $P + \Pi$ near the critical point. The net baryon distribution is also modified at forward rapidity, but is caused only by the enhanced convection owing to the reduced effective pressure because net baryon number is conserved and is not directly affected by the entropy production.

\begin{figure}[tb]
\center
\includegraphics[width=2.5in]{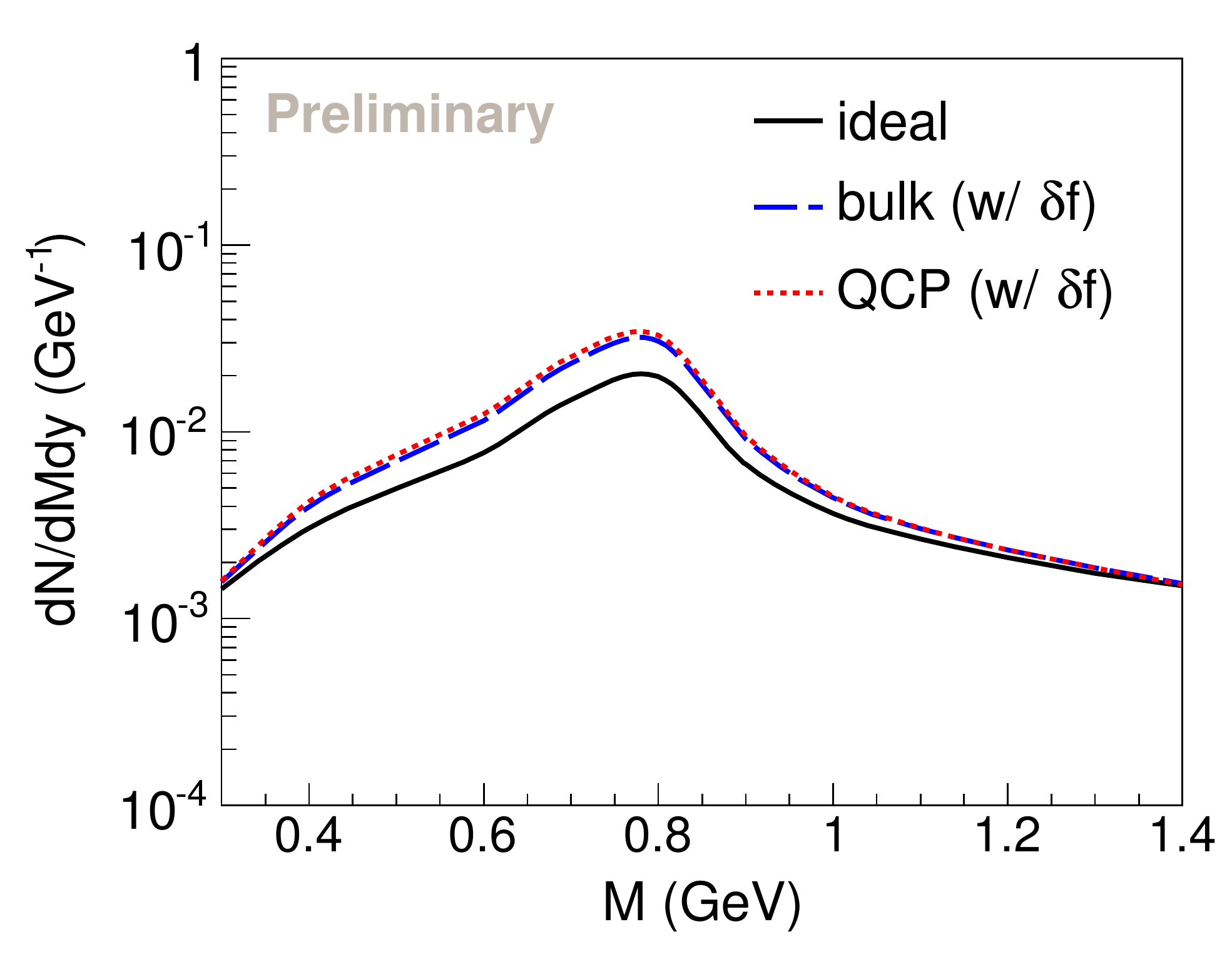}
\includegraphics[width=2.5in]{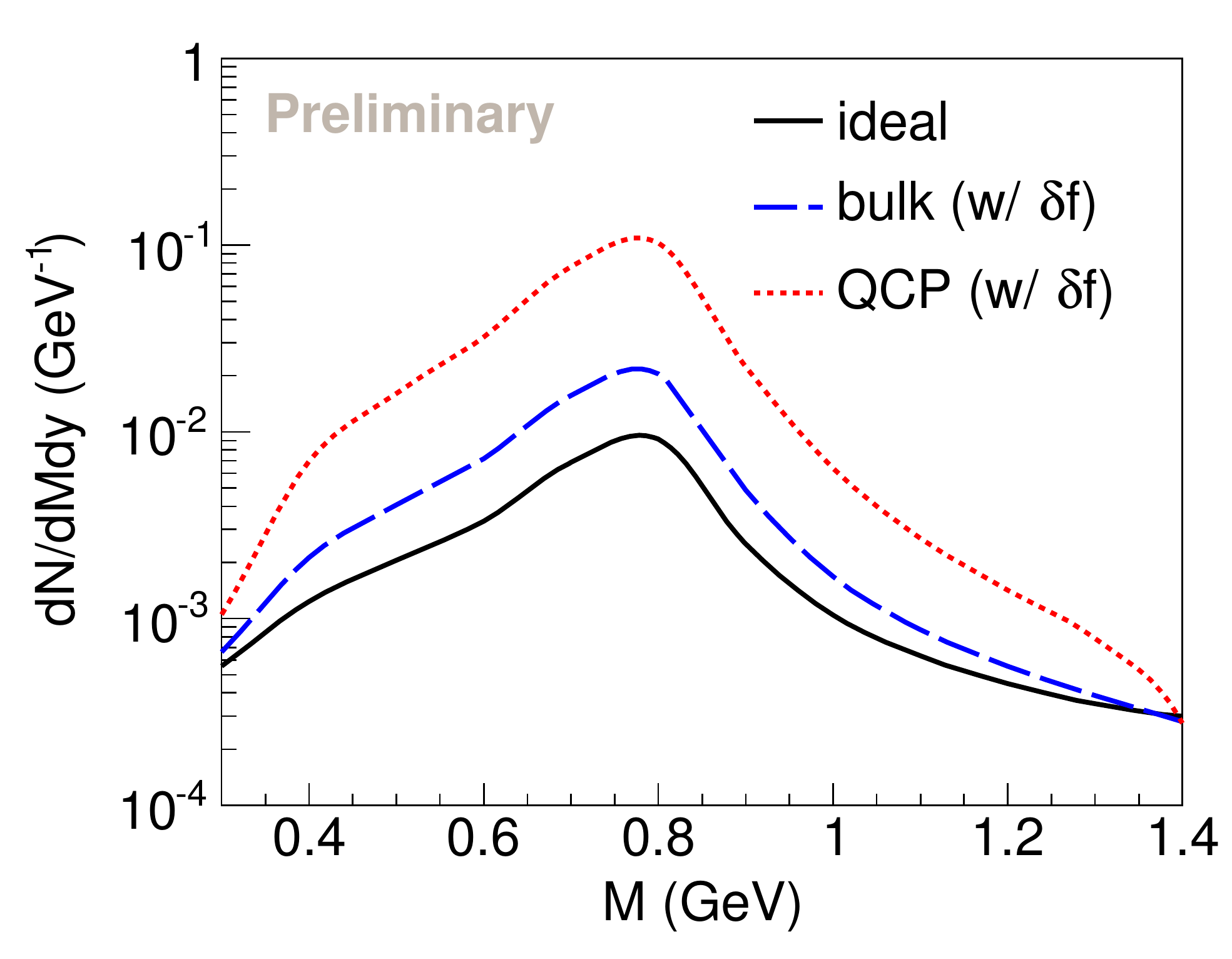}
\caption{Invariant mass spectra of thermal dileptons at (left) $Y=0$ and (right) $Y=2$.}
\label{fig:3}
\end{figure}

We further study the invariant mass spectra of thermal dileptons (Fig.~\ref{fig:3}). One can see that bulk viscosity enhances the spectra at lower momentum because the lifetime of the fireball is longer in an off-equilibrium system. The QCD critical point does not have effect at $Y=0$, but could noticeably modify the spectra at $Y=2$ where the fluid elements pass through the critical region. It should be noted that since we consider a simplified pion gas model only the peak coming from $\pi^+ \pi^- \to \rho \to l^+ l^-$ is seen. It remains to be seen how the full hadronic processes would appear in the spectra with bulk viscosity .

\section{Summary and outlook}
\label{sec:4}

We have developed a bulk viscous hydrodynamic model near the QCD critical point based on the observation that critical enhancement of the transport coefficients can occur $\zeta \sim \tau_\Pi \sim \xi^3$. Numerical demonstrations indicate that the critical point can enhance rapidity distributions of charged hadrons through the viscous entropy production and the extra convection caused by the gradient in $P+\Pi$. The latter process also enhances the net baryon distribution. Dilepton invariant mass spectra can be affected by the critical point through the viscous distortion of the phase-space distribution. Future prospects include derivation of the full hadronic dilepton emission rate, comparison of different $\delta f$ models and estimation of radial flow effects.

\section*{Acknowledgments}
We would like to thank G.~Denicol, U.~Heinz, R.~Pisarski, L.~McLerran, K.~Rajagopal,
P.~Sorensen, and M.~Stephanov for very valuable discussions and M.~Nahrgang, T.~Sch\"{a}fer, and
B.~Schenke for commenting on the manuscript. AM was supported in part by JSPS Overseas Research Fellowships.
This material is based upon work supported by the U.S. Department of Energy, Office
of Science, Office of Nuclear Physics, under Contract No. DE-SC0012704, and within
the framework of the Beam Energy Scan Theory (BEST) Topical Collaboration.

\bibliographystyle{elsarticle-num}
\bibliography{<your-bib-database>}


\end{document}